\documentclass[a4paper,11pt]{article}
\usepackage{upgreek}
\usepackage[]{graphicx}
\usepackage[]{xcolor}
\usepackage[margin=1in]{geometry}

\title{Search for jet quenching in small systems}
\date{}

\author{Filip Krizek\footnote{Nuclear Physics Institute of the Czech Academy of Sciences,\\ \textcolor{white}{.....} Hlavni 130, Husinec--Rez, Czechia \vadjust{\vskip 1mm \vskip 0pt}\\   \textcolor{white}{.....} E-mail: krizek@ujf.cas.cz
}\quad for the ALICE Collaboration}

\begin{document}
\maketitle
\setcounter{secnumdepth}{0}

\section{Abstract}
High multiplicity final states of small collision systems, such as proton--proton or proton--nucleus,
exhibit some signatures which  resemble features associated with QGP formation in heavy-ion collisions, 
e.g., collective phenomena or enhancement in produced strangeness. At the same time, 
there is no experimental evidence for QGP-induced jet quenching to date.
Thus, quantification or setting limits on the magnitude of jet quenching in small systems is essential 
for understanding the conditions needed for QGP formation. These proceedings discuss several recent measurements
that searched for jet quenching effects in small collision systems.

\section{Search for jet quenching in small systems}
Quark--Gluon Plasma (QGP) is a phase of strongly interacting matter, 
which consists of deconfined quarks and gluons \cite{QGP}.
High energy densities and temperatures, necessary to produce the QGP, can be reached 
by colliding ultrarelativistic heavy nuclei. Such collisions produce a finite volume of this medium.
The experimental signatures, which are expected to emerge due to QGP formation, include  
strong collective behavior of final state particles, 
enhanced production of strange particles, and jet quenching.
Jet quenching was observed for the first time at the RHIC experiments \cite{STAR, PHENIX}.
It was manifested as a large suppression in the yield of high transverse momentum ($p_{\rm T}$) 
hadrons measured in nucleus--nucleus collisions, 
compared to the yield expected from the corresponding number of nucleon--nucleon collisions, 
approximated by a scaled pp reference.
This ratio is known as the nuclear modification factor.
Jet quenching arises from interactions between partons and the created QGP medium.
These interactions redistribute the energy of the partonic shower and cause, in addition to energy loss, 
jet substructure modification, and medium-induced dijet acoplanarity \cite{Acopl}.

Collectivity and strangeness enhancement were also observed in much smaller collision systems
having just a few interacting nucleons, such as p--Pb or pp.
The fact that there is a lack of evidence for jet quenching in small collision systems, raises the question of 
whether the QGP formation is the only possible source that can cause collectivity and strangeness enhancement. 
The size of jet quenching is expected to be small in small collision systems \cite{Tywo,OO}, 
which poses a challenge for the experiment. 
Let us illustrate this with the example of p--Pb collisions. 
Minimum bias p--Pb collisions are dominated by events with a large impact parameter,
where the size of the produced hot zone is small.
Hence the path length traversed by a parton in this zone is expected to be small on average, 
leading to a small jet quenching effect on average.
The produced volume can be increased by imposing a bias on the minimum multiplicity of produced particles,
as the multiplicity is expected to be correlated with collision-zone volume size on average.
The problem is that in this case the Glauber model 
provides the number of binary nucleon--nucleon collisions
 with considerable uncertainty because of multiplicity
fluctuations \cite{OO}.
In contrast, for MB collisions, the pp reference scaling factor can be calculated analytically.

\begin{figure}[htbp]
\centering
	 \includegraphics[width=0.5\textwidth]{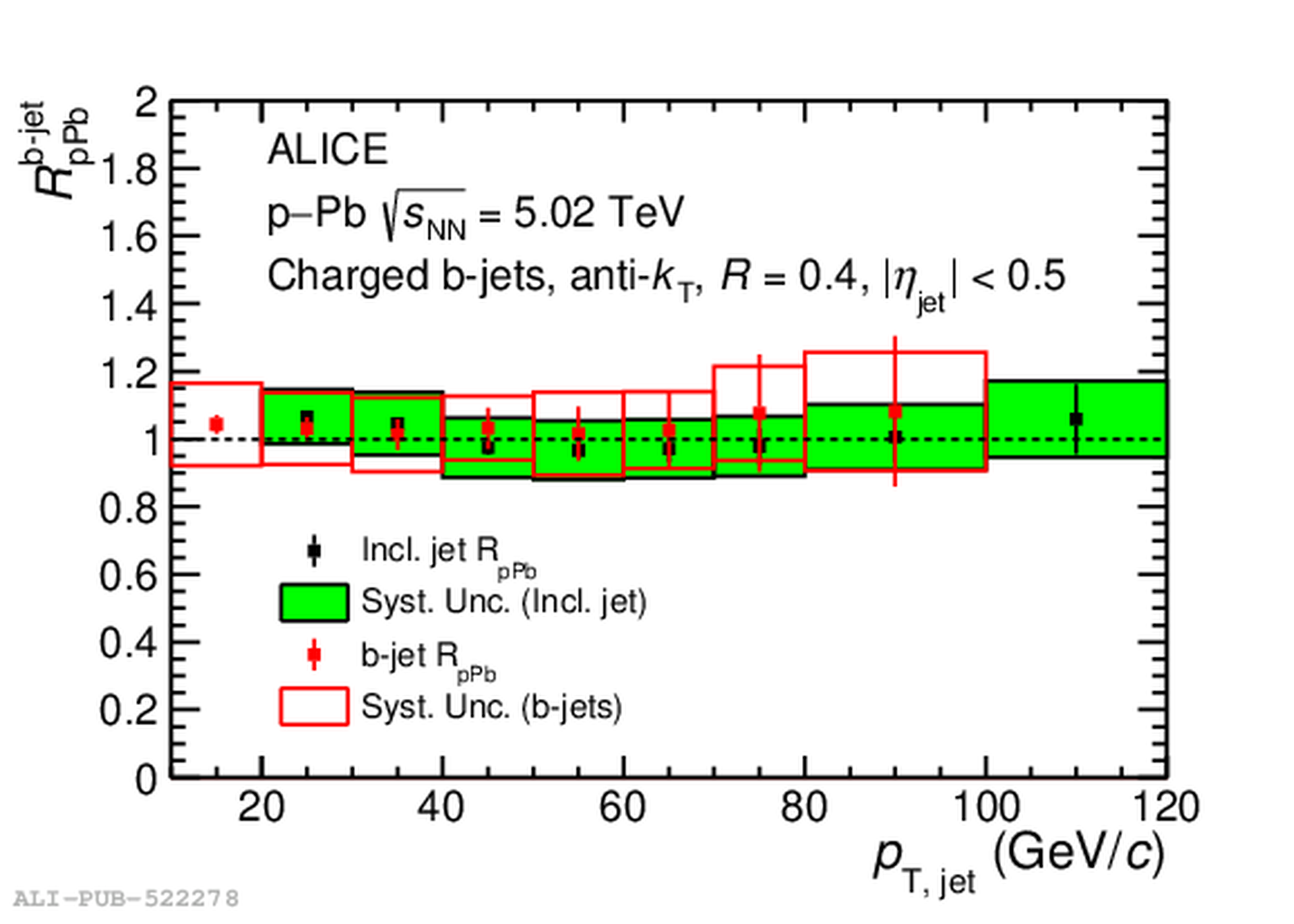}
	 \caption{Nuclear modification factor for inclusive  charged-particle b jets and inclusive charged-particle  jets in p--Pb collisions at $\sqrt{s_{\rm NN}}=5.02$ TeV.
	 Jets were reconstructed using the anti-$k_{\rm T}$ algorithm \cite{akt} with resolution parameter $R=0.4$. 
	 Their pseudorapidity was constrained to $|\eta_{\rm jet}| < 0.5$. Taken from Ref.~\cite{ALICEbjet}.	 
	 }
\label{figBjet}       
\end{figure}

Figure \ref{figBjet} shows the nuclear modification factor $R_{\rm pPb}$ 
for inclusive charged-particle b jets and inclusive charged-particle jets 
in minimum bias p--Pb collisions at $\sqrt{s}=5.02$ TeV \cite{ALICEbjet}. The data are compatible with unity.
This means that possible modification due to jet quenching, if it exists, is below the resolution of this measurement.
The measurement similarly does not exhibit any sign of quark-mass-dependent effects.

%
%

\begin{figure}[htbp]
\centering
	 \includegraphics[width=1.\textwidth]{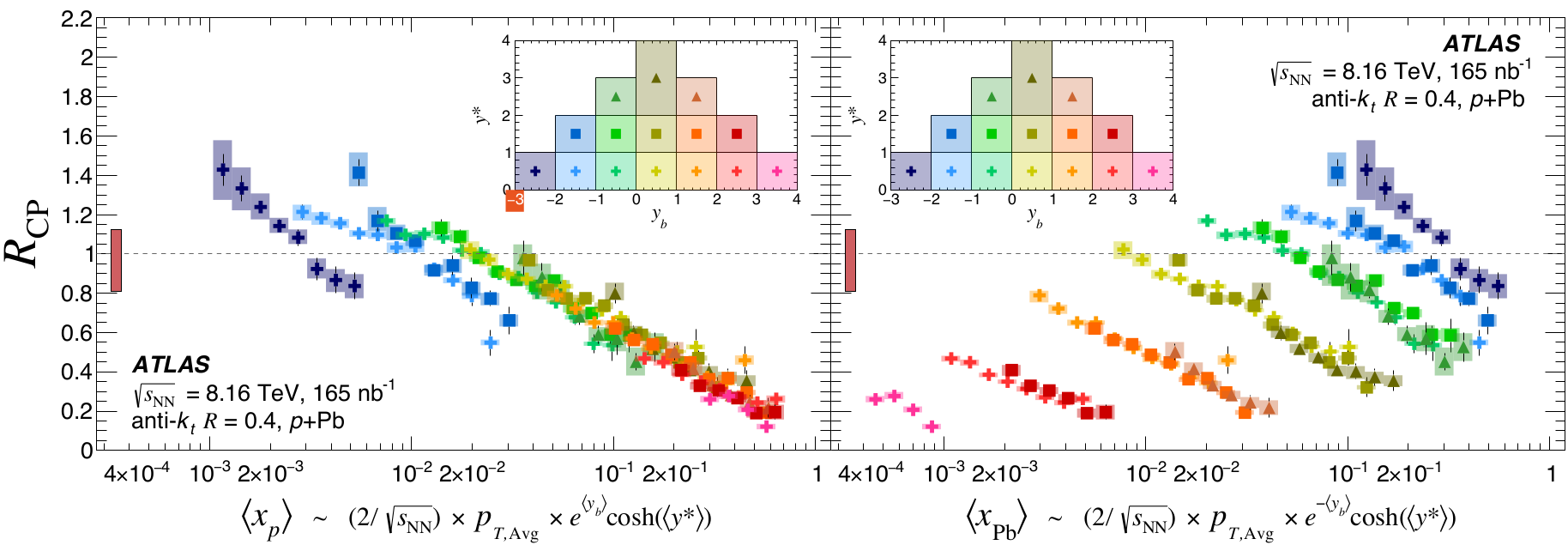}
	 \caption{Dijet $R_{\rm CP}$ as a function of the mean proton-energy fraction carried by the proton-projectile (left)
	 and Pb-target parton (right) in p--Pb collisions at $\sqrt{s_{\rm NN}}=8.16$ TeV as measured by ATLAS.
	 Color markers correspond to different configurations of dijet boost $y_{\rm b}$ and half-rapidity separation $y^{*}$. 	 
	 Figure from Ref.~\cite{ATLASdijet}.	 
	 }
\label{figRCPATLAS}       
\end{figure}

The nuclear modification factor of high-$p_{\rm T}$ jets in event-activity-biased 
p--Pb collisions was reported by ATLAS \cite{ATLASRpPb}.
The event activity was determined based on the energy registered 
in a forward calorimeter placed in the Pb-going direction.
The results exhibit a striking dependence of the $R_{\rm pPb}$ on the pseudorapidity of produced jets.
While in the Pb-ion-going direction, the $R_{\rm pPb}$ is compatible with unity,
the jets produced in the proton-going direction exhibit 
suppression in $R_{\rm pPb}$ for central events and enhancement for peripheral events. 
At the same time, the minimum bias data are compatible with unity.
To understand this behavior ATLAS carried out a new measurement \cite{ATLASdijet}, 
where they investigated how the kinematics of the initial hard parton scattering process
 affects the jet yields in central p--Pb relative to peripheral collisions.
The average momentum fractions of the interacting partons in the proton projectile $\left<x_{\rm p}\right>$ and 
Pb target $\left<x_{\rm Pb}\right>$ can be inferred from the properties of produced dijets and expressed
in terms of dijet boost $y_{\rm b}$, half-rapidity separation of the corresponding jets $y^{*}$, 
and their average transverse momentum $p_{\rm T,Avg}$. Figure~\ref{figRCPATLAS} shows
the ratio of the yield measured in central collision divided by the binary-collision-scaled
yield from peripheral collisions ($R_{\rm CP}$) 
as a function of the fraction of momentum carried by a parton in the proton projectile
and a fraction of momentum carried by a parton in the lead target.
The tilt observed in data reveals that scattering of large-$x$ partons
will enhance the fraction of peripheral events. 
This is qualitatively compatible with a picture
where a proton with a large-$x$ parton will have a smaller size
as less energy is available for color charge fluctuations \cite{Fluct}.
The data further exhibit pronounced scaling behavior in the valence region of the projectile proton. 
On the other hand, the $R_{\rm CP}$ does not exhibit such scaling as a function of the momentum fraction 
carried by the target-nucleus parton. 

Semi-inclusive observables, which include coincidence measurements of trigger hadrons and jets 
with another object (a hadron or jet) in a given event, can be used to search for jet quenching.
In this case, the basic observable is the per-trigger-normalized yield of associated objects.
As the yield of triggers and associated objects is affected by the same initial geometry conditions,
these observables do not require Glauber model scaling and can be compared directly to a pp reference.

\begin{figure}[htbp]
\centering
	 \includegraphics[width=0.9\textwidth]{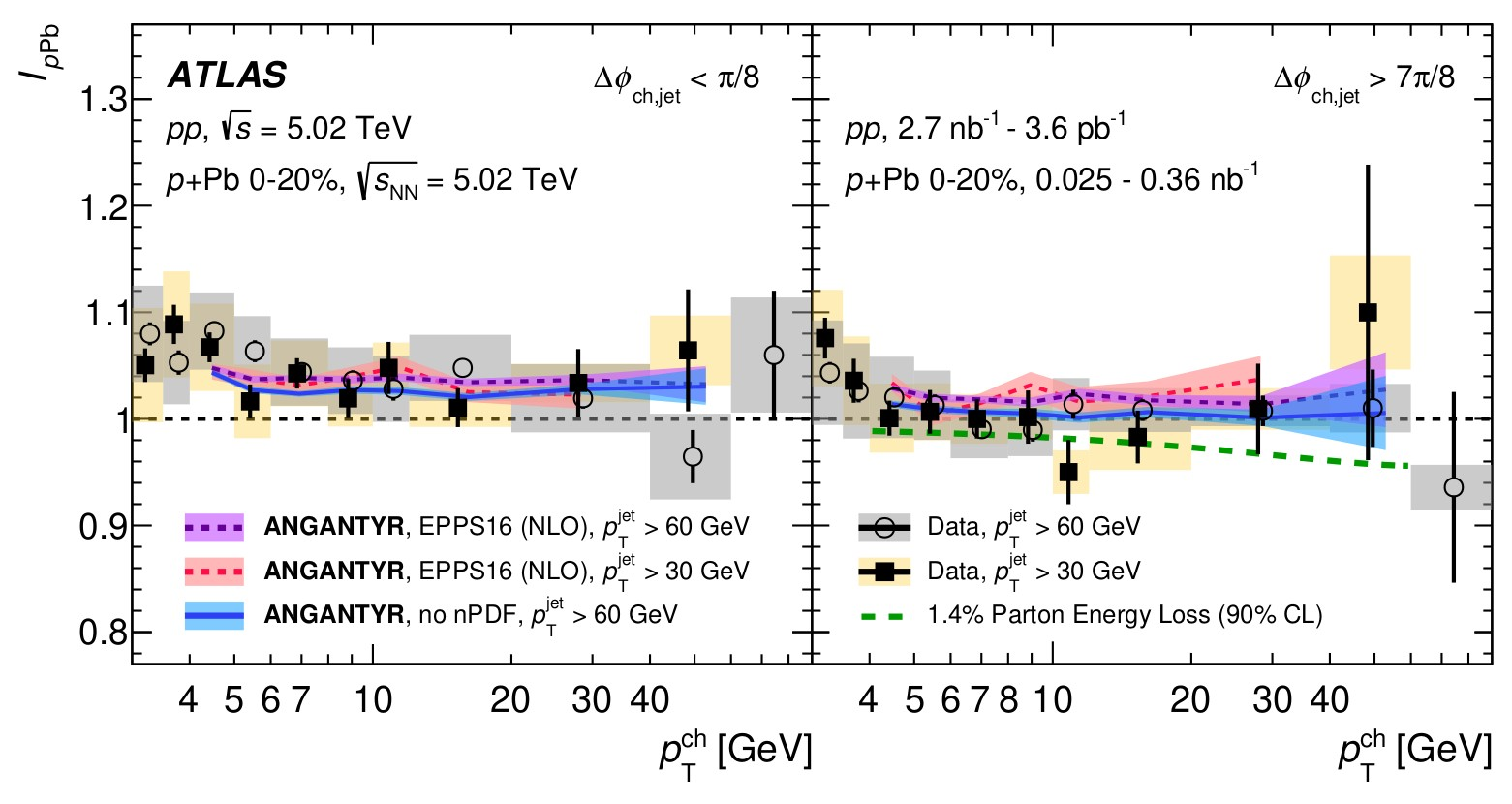}
	 \caption{ $I_{\rm pPb}$ of  charged-hadrons associated with a high-$p_{\rm T}$ jet trigger  
	 ( $p_{\rm T}^{\rm jet} > 60$ GeV/$c$ and $p_{\rm T}^{\rm jet} > 30$ GeV/$c$)
         in p--Pb collisions at $\sqrt{s_{\rm NN}}=5.02$ TeV as measured by ATLAS.
         The $I_{\rm pPb}$ is presented separately for near-side (left) and away-side (right) hadrons
	 as a function of associated charged-hadron $p_{\rm T}$. 
	 Data are compared with PYTHIA 8 Angantyr (bands). 
	 The 90\% CL limit on the potential jet-quenching energy loss is illustrated by the green dashed line.
	 Figure from Ref.~\cite{ATLASjeth}. 	 
	 }
\label{figIpPbATLAS}       
\end{figure}

ATLAS has recently reported a measurement, where they set a strong constraint on jet quenching \cite{ATLASjeth}. 
The analysis used events that were tagged by the presence of a trigger jet having $p_{\rm T}^{\rm jet} > 60$ GeV/$c$ 
or $p_{\rm T}^{\rm jet} > 30$ GeV/$c$. 
In these events, they analyzed jet--hadron correlations in the near-side and away-side regions,
which were defined using azimuthal-angle separation between the trigger-jet axis and associated hadron;
for the near-side region the separation was less than $\pi/8$ rad and for the away-side region it was larger than $7\pi/8$ rad.
Event activity was defined based on a signal provided by neutron spectators in a Zero Degree Calorimeter.
The ratio of the per-trigger-normalized charged-hadron yields measured in p--Pb and pp collisions ($I_{\rm pPb}$) 
 is shown as a function of hadron $p_{\rm T}$ in Fig.~\ref{figIpPbATLAS}. 
The data suggest that the $I_{\rm pPb}$ of away-side hadrons is compatible with unity while at the near-side, 
 $I_{\rm pPb}$  exceeds unity.
The data are compatible with a prediction by Angantyr \cite{Angantyr}, which is a PYTHIA-based model for 
heavy-ion collisions that does not account for jet quenching. 
Moreover, ATLAS verified that the near-side enhancement is not due to PDF and isospin effects.
The data on the away side were further compared with a PYTHIA-based energy-loss model.
It was found that the parton opposite to a $p_{\rm T}^{\rm jet} > 60$ GeV/$c$ jet trigger
 loses less than 1.4\% (at 90\% CL) of its energy before undergoing vacuum-like fragmentation into
$p_{\rm T} > 4$ GeV/$c$ charged hadrons.

\begin{figure}[htbp]
\centering
	 \includegraphics[width=0.65\textwidth]{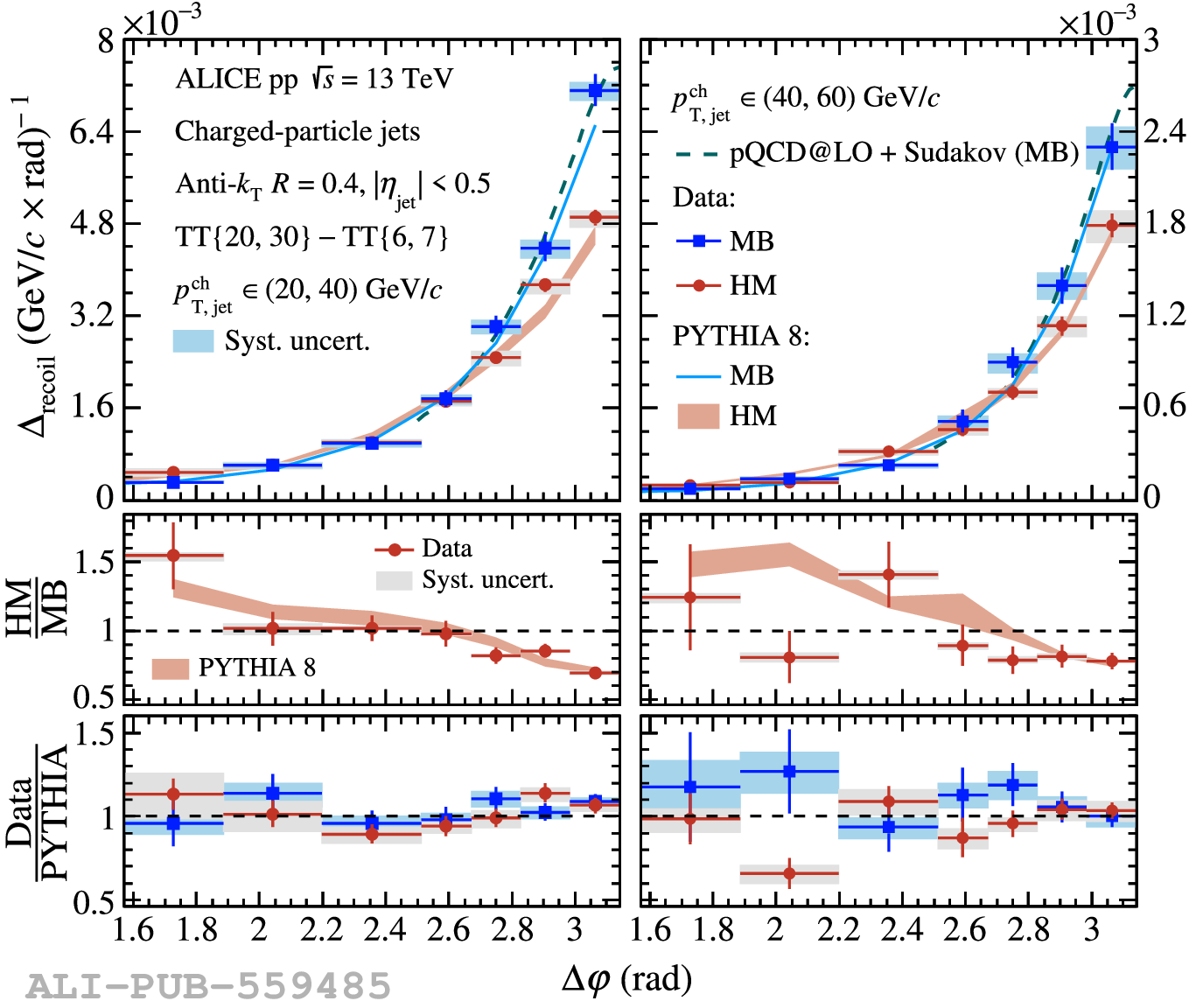}
	 \caption{Top panel: $\Delta_{\rm recoil}$ distributions measured in HM and MB pp collisions at $\sqrt{s}=13$ TeV.   
		   Recoil jets were reconstructed from charged hadrons using the anti-$k_{\rm T}$ algorithm
		   with resolution parameter $R=0.4$. Jets were associated with trigger hadrons with 
		  transverse momentum constrained to 20--30 GeV/$c$ and 6--7 GeV/$c$.
		  The transverse momentum of recoil jets is constrained to 20--40 GeV/$c$ (left) and 40--60 GeV/$c$ (right).
		  The HM and MB $\Delta_{\rm recoil}$ distributions calculated by PYTHIA 8 Monash \cite{Monash} 
		  are represented by color bands.
		  Middle panel: Ratio of HM and MB distributions.
		  Bottom panel: Ratio of data to PYTHIA.
		  Figure from Ref.~\cite{ALICEhjet}. 
	 }
\label{figDrecoil}       
\end{figure}

\begin{figure}[htbp]
\centering
	 \includegraphics[width=0.67\textwidth]{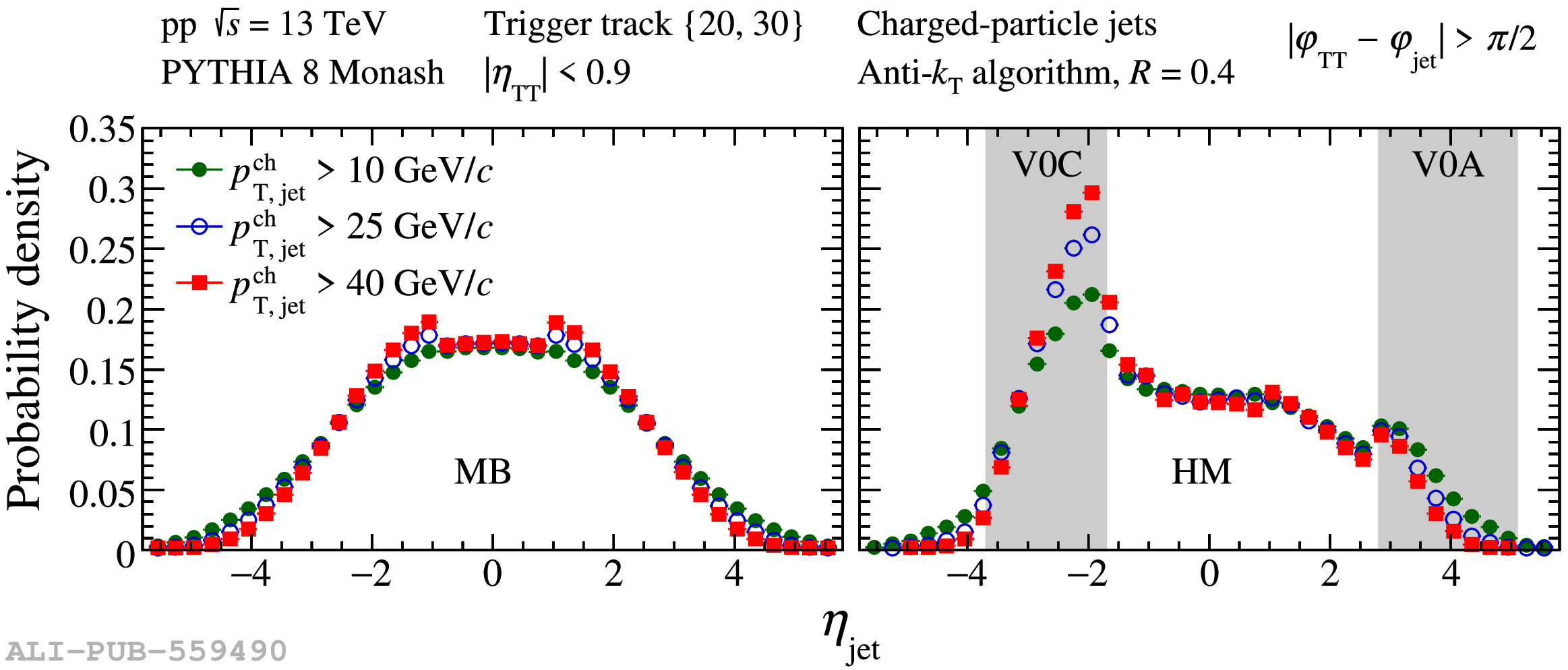}
	 \caption{Pseudorapidity distribution of jets recoiling from a 20--30 GeV/$c$ 
	          hadron trigger in minimum bias (left) and high-multiplicity (right) 
		  proton--proton collisions at $\sqrt{s}=13.6$ TeV simulated by PYTHIA 8 Monash \cite{Monash}.
		  The coverage of the V0 detector is highlighted in grey.
		  Figure from Ref.~\cite{ALICEhjet}. 
	 }
\label{figPseudo}       
\end{figure}

The semi-inclusive approach was also used by the ALICE collaboration 
to search for jet quenching in high-multiplicity (HM) pp collisions \cite{ALICEhjet}.
The selected HM events had more than 5 times larger charged-particle multiplicity 
in the forward V0 scintillator arrays compared to minimum bias (MB) events. 
This sample represents 0.1\% of the MB cross section.
ALICE analyzed the distribution of 
the azimuthal angle $\Delta\varphi$ between a high-$p_{\rm T}$ trigger hadron 
and associated recoiling jets to look for medium-induced broadening of the distribution in HM events.
The analysis utilized a data-driven method to remove the contribution of jets uncorrelated
with the trigger hadron.
The jet yield which is correlated with the trigger hadron, was obtained by subtracting 
 two per-trigger normalized distributions that are associated with two exclusive trigger $p_{\rm T}$ bins.
The corresponding subtracted yield is called $\Delta_{\rm recoil}$.  The fully corrected
$\Delta_{\rm recoil}$ distributions are shown for HM and MB events as a function of $\Delta\varphi$ in Fig.~\ref{figDrecoil}.
The HM distributions are suppressed w.r.t. MB around $\Delta\varphi = \pi$, which resembles a jet quenching effect,
however, PYTHIA, which does not account for jet quenching, reproduces the suppression.
The PYTHIA generator was therefore used to investigate the origin of this phenomenon.
Figure \ref{figPseudo} presents pseudorapidity distributions of  high-$p_{\rm T}$ recoil jets 
in MB and HM events simulated by PYTHIA. In the case of MB events, the distributions are symmetric around midrapidity,
while in HM events, recoil jets are biased to appear in the acceptance of the V0 arrays.
These recoil jets are thus missing in the central barrel to balance the high-$p_{\rm T}$ trigger hadron.
This measurement demonstrated that jet-quenching effects can be masked by effects stemming from
event selection biases.

In summary, there is no confirmed evidence of jet quenching effect in small collision systems. 
The magnitude of such an effect is likely to be small. Its observation in an experiment is
complicated because jet quenching signatures can be masked by event selection biases, 
such as picking up fluctuations in the proton wavefunction enhanced by an imposed event activity bias or
emergence of NLO processes. 
Attribution of the non-zero  elliptic flow of high-$p_{\rm T}$ hadrons 
 observed in p--Pb collisions \cite{ATLAShighv2}  to jet quenching remains an open question.

F. Krizek acknowledges support by the project OPJAK CZ.02.01.01/00/22\_008/0004632.

\end{document}